\documentclass[aps,prb,twocolumn,groupedaddress,showpacs]{revtex4}

\usepackage{graphicx}

\begin{document}

\title{Inhibition of protein crystallization by evolutionary negative design}

\author{Jonathan P.~K.~Doye}
\email[]{jpkd1@cam.ac.uk}
\affiliation{University Chemical Laboratory, Lensfield Road, Cambridge CB2 1EW, United Kingdom} 
\author{Ard A. Louis}
\email[]{aal20@cam.ac.uk}
\affiliation{University Chemical Laboratory, Lensfield Road, Cambridge CB2 1EW, United Kingdom} 
\author{Michele Vendruscolo}
\email[]{mv245@cam.ac.uk}
\affiliation{University Chemical Laboratory, Lensfield Road, Cambridge CB2 1EW, United Kingdom} 

\date{\today}

\begin{abstract}
In this perspective we address the question: 
why are proteins seemingly so hard to crystallize? 
We suggest that this is because of evolutionary negative design, 
i.e.\ proteins have evolved not to crystallize, because crystallization, 
as with any type of protein aggregation, compromises the viability of the cell.
There is much evidence in the literature that supports this hypothesis, 
including the effect of mutations on the crystallizability of a protein, 
the correlations found in the properties of crystal contacts in 
bioinformatics databases and the positive use of protein crystallization 
by bacteria and viruses.
\end{abstract}

\pacs{87.15Nn,87.14Ee}

\maketitle

The overwhelming impression one gets from reading the literature
on protein crystallization and listening to 
experts is that protein crystallization is difficult
and requires considerable effort. Furthermore,
experience and a certain feeling for what might work can play a crucial role.
Recent technical innovations,\cite{Bergfors}
such as the availability of 
scanning kits which codify experience to scour for appropriate
crystallization conditions, have helped to 
provide valuable savings in labour. 
These advances, however,
have not altered what seems to be the basic fact:
Proteins, for the most part, 
do not seem to want to crystallize, and have to be coaxed into doing so
through the use of suitable cunning.

This situation is particularly vexing, 
because protein crystallization is a vital step in protein structure
determination, and hence to structural genomics initiatives,\cite{Burley99} 
which seek to catalogue the protein structures associated with 
the whole genome of a target organism.
Although there are also obstacles associated with the expression
and purification of the proteins, crystallization is often labelled
as the major bottleneck in this process.\cite{Vekilov02}

The quantification of some of the difficulties involved
in protein crystallization
is beginning to emerge from structural genomics pilot studies.
Generally, the output of new protein structures so far has been
``disappointingly low''.\cite{Service02}
For example, for a thermophilic prokaryote, probably the class of
organisms for which the greatest success rate is expected, only 13\% of
a target set of non-membrane proteins were estimated to be readily
amenable to structural determination; at present only 4\% of the structures of
these proteins have actually been obtained.\cite{Yee03}
These successes probably represent the ``low-hanging fruits'' of the
proteome. How to reach higher branches remains unclear.

In this perspective, we would like to take a step back and 
offer our opinions on an important question raised by this situation: 
Why is the crystallization of proteins so difficult? 
This is not only a fundamental question, but also a practical one. 
A natural starting point for any rational attempt to overcome the 
obstacles that hinder protein crystallization is to first 
understand the nature of these barriers.

In general, one expects
that it should be possible to obtain crystals for soluble 
molecules that have a well-defined structure.\cite{glass} 
So why should globular proteins be any different? 
One possible answer is that proteins are polypeptide chains 
with significant conformational entropy 
and this will have some effect on their crystallization properties.
However, their dynamic nature does not interfere with their 
ability to form specific complexes with proteins and other molecules. 

In our opinion, the answer to this question lies in the 
evolutionary origin of proteins.  Proteins are a very special type of
polymer and their possible states are different from those of normal
polymers.  For example, simple homopolymers can be either in a swollen
or a collapsed phase, depending on the quality of the
solvent.\cite{deGennes79}  But whereas proteins in a collapsed
globular state can remain soluble for appreciable concentrations,
collapsed homopolymers aggregate very easily. There are, of course,
many more differences between simple polymers and proteins. Here we
suggest that evolution appears to have enhanced the tendency to
keep globular proteins soluble and active, reducing the probability of
realizing all types of aggregate states.

Our hypothesis is thus that proteins have evolved
not to crystallize, because crystallization, 
as well as any type of aggregation, compromises the viability of the cell.
Most aggregation diseases, e.g. Alzheimer's and Creutzfeldt-Jakob
disease, are associated with non-native protein structures, and the
cell has developed sophisticated quality control mechanisms to cope with 
misfolded proteins.\cite{Kopito00}  
However, there are also a number of diseases
associated with the aggregation of proteins in their native state.
Perhaps the best known example is sickle cell anaemia, where a mutant form of
hemoglobin coalesces to form ordered fibrillar aggregates inside red
blood cells. In addition, there are also instances of diseases that result from
crystallization: Certain forms of cataracts and anaemia are caused by
crystallization of mutant forms of the $\gamma$
crystallin\cite{Pande01} and hemoglobin\cite{Vekilov02b} proteins,
respectively.  Furthermore, protein crystallization has been found to
be associated with other pathologies\cite{McPherson99}. In
general, however, such diseases are less common that those associated
with the aggregation of misfolded proteins. 
We suggest that this difference 
is because the well-defined structure of the native state 
makes it much more amenable to evolutionary control.

One further consideration is that the selection pressure is with respect to 
crystallization {\it in vivo},
whereas protein crystallographers explore far-from-physiological
conditions {\it in vitro}.  However, in our view, the fact that crystallization
is difficult even in the latter circumstances simply reflects the
robustness of the strategies used by nature to ensure that proteins do not
crystallize in the cellular environment.

Our hypothesis is one example of a negative design principle. More often
we think in terms of positive design, i.e.\ that the sequence of a protein
has been optimized through evolution to give the protein particular 
characteristics.
However, negative design leading to the avoidance of unwanted properties,
such as crystallizability or aggregation, can be equally important. 

Such negative design principles have been previously proposed for 
both the single-molecule and intermolecular properties of proteins. 
For example, for a protein to fold reliably to its native state, 
not only must the native structure be particularly low in free energy, 
but alternative conformations must also not have similar
or lower stability.\cite{Hellinga97}
Some of the strategies by which this specificity can be achieved have been
identified and then applied in the {\it de novo} design of 
proteins.\cite{DeGrado99} For example, even though it is generally more
thermodynamically favourable to have hydrophobic residues in the core of the
protein, greater specificity can be achieved by the introduction 
of some interacting polar residues into the core.\cite{Bolon01}

Lessons on negative design can be learnt from the necessity
to avoid aggregation. This is a particular problem
for proteins involving $\beta$-sheets, since their edges are natural sites
for association with other $\beta$-sheets in nearby proteins, 
and, for example,  can lead to the 
extended $\beta$-sheet structures found in amyloid deposits.
A number of negative design strategies have been found in 
natural proteins that protect $\beta$-sheet edges.\cite{Richardson02} 
The simplest strategy is to form a continuous $\beta$-sheet structure without 
any edges, as in $\beta$-barrels.
Another of the identified strategies has been successfully applied to 
turn an aggregating protein into a soluble monomeric form 
by a single mutation of a non-polar residue to lysine.\cite{Wang02b}

Designing out unwanted interactions is also necessary in 
molecular recognition. To achieve specificity, a protein must not only 
interact strongly with the target molecule, but also 
have much less favourable interactions with all other 
molecules.\cite{Havranek03,Shifman03}

The two examples discussed above illustrate the combination of positive and 
negative design that is used to tailor the interprotein interactions.
Most generally, this is seen in 
the remarkable properties of cellular solutions, where
crowded, multi-component mixtures with protein packing fractions of 
up to 40\%\cite{Ellis03}
can be both functionally active and stable.
By contrast, any attempt to make artificial nanocolloidal 
mixtures of similar density is bound to result in components sticking together
to form an amorphous deposit.
In fact, colloid scientists expend considerable
effort modifying the surfaces of  colloids---adding, for example, charged
groups or short polymer brushes---to prevent this from occurring.
To achieve this combination of specific attraction (positive design)
and generic repulsion (negative design), evolution must exert remarkable
control over the matrix of all possible interprotein interactions.\cite{Sear03b,Searunpub}
In this context, our hypothesis concerns a particular type
of interaction (namely crystal-forming) that contributes to the
diagonal elements (i.e. self-interactions) of this matrix.

Let us consider how this negative design might be achieved.
As many amino acid sequences can give rise to the same final
protein fold, there is considerably freedom in how the 
amino acids, particularly those on the surface of the protein,\cite{Bastolla03} 
are chosen. 
This flexibility could potentially allow the protein surface 
to be organized such that crystallization is hindered, 
without affecting either the structure of the 
protein's fold or its active site.

Importantly, such a scenario has testable consequences. 
If
the surfaces of proteins have
been optimized to sufficiently reduce their crystallizability, one
would expect that random mutations of the surface amino acids that 
do not alter the structure of the protein fold or its activity (i.e.\ 
only the `neutral' mutations that are evolutionarily allowed)
would be likely to lead to a more crystallizable protein.
By contrast, if our hypothesis did not apply and 
a protein's crystallizability did not influence the choice of 
surface amino acids, 
one would expect such mutants to be as likely to hinder as
to enhance a protein's crystallizability.

We know of two such systematic 
studies of the crystallizability of mutagens,
the first on human thymidylate synthase\cite{McElroy92} and the
second on a fragment of the DNA gyrase B subunit from 
{\it Escherichia Coli}.\cite{Darcy99}
In both studies, mutations were found to have a dramatic effect
on the crystallization properties of the protein. 
In agreement with our negative design hypothesis, the mutants generally 
showed enhanced crystallizability
compared to the wild-type, as measured by the number of hits in a 
crystallization screen. There was also evidence of enhancement in
crystal quality. Moreover, some of the mutants crystallized in space groups
that were not encountered for the wild-type protein.
Although the amount of data is not enough to provide conclusive
justification of our negative design argument, it is strongly suggestive.
Furthermore, there is a body of more anecdotal evidence consistent 
with our ideas, namely the growing catalogue
of proteins that have been first crystallized as mutants.\cite{Dale03} 

By contrast, 
where there has been positive design of the protein surface, 
as in the case of specific functional binding interactions between two proteins,
one would expect random mutagenesis to lead on average 
to a reduction in the binding affinity between the proteins. 
This is indeed the case, and such studies have 
played an important role in understanding the nature of protein-protein 
binding through the identification of small sets of residues that 
are key to the stability of the interface.\cite{Bogan98}

Although it seems clear that the surfaces of proteins have been
designed to hinder crystallization,
there still remains the question of what physical mechanism underlies
the reduced crystallizability of the evolutionary selected protein
surfaces.  One might guess that this behaviour reflects some complex
property of the surface, and hence would be hard to identify or
rationally control.
However, there is experimental evidence that surface lysine residues 
could play a key role in this negative design strategy. 

As one would
expect for a charged amino acid, lysine prefers to be at the surface
of the protein, where it can interact with the aqueous environment.
In fact, lysine has the highest propensity to be at the surface of all 
the amino acids and is the most common surface residue.\cite{Baud99}
Lysine is also unique in presenting
the largest amount of solvent accessible surface area that is hydrophobic in 
character,\cite{Lins03} because of the long hydrophobic tail that 
links the amine group to the protein backbone. Even more interestingly
for our present considerations, systematic studies of interprotein contacts
have found lysine to be the most underrepresented 
amino acid at crystal contacts,\cite{Dasgupta97,Iyer00} 
and even more so at the interfaces between subunits of protein 
oligomers\cite{Dasgupta97} and between proteins that form functional 
complexes.\cite{Jones96,LoConte99}
These negative correlations of course raise questions concerning
the purpose of lysine residues: Why are they so abundant on the surface, 
if they are only reluctantly involved in functional interactions?
It could be that lysine plays an important negative role in regulating 
interprotein interactions through preventing unwanted interactions.
Indeed, Dasgupta {\it et al.} suggested
the mutation of lysine residues as a rational strategy for enhancing
crystallizability.\cite{Dasgupta97} 

Just such an approach has been implemented in the experiments of the 
Derewenda group.\cite{Longenecker01,Longenecker01b,Garrard01,Mateja02}
They considered the effects of a series of lysine to alanine mutations
for human RhoGDI.\cite{Longenecker01}
Their rationale for this particular type of mutation 
was that the substitution of an amino acid with high conformational 
entropy by a smaller one would lead to a reduction in the 
entropy loss on crystal contact formation. 
Whether for this reason or not---the replacement of a charged amino acid by 
a neutral one will also lead to concomitant changes in the 
electrostatic interactions---the results were dramatic.
The mutants invariably showed enhanced crystallizability, and
often produced crystals that diffracted to higher resolution than
achievable otherwise.
Consistent with the idea that the lysine residues somehow prevent
unwanted interactions, new crystal contacts were often formed at 
the sites of the mutations. 
A similar study on glutamate to alanine mutations also 
revealed enhanced crystallizability, 
although not quite to the same degree.\cite{Mateja02} 
This rational mutagenesis strategy has since been successfully 
applied to crystallize proteins of previously 
unknown structure.\cite{Longenecker01b,Garrard01}

Additional support for the idea that negative design is
a key aspect of evolution at the molecular level comes from instances where
one of the assumptions of our hypothesis does not hold; namely, 
that crystallization is harmful to the cell. 
Although this assumption is likely to be generally true, it is
a simplification and will not necessarily hold for all cellular 
environments. In the absence of such a selection pressure, crystallization
is likely to be significantly easier. 
Indeed, there may even be circumstances when crystallization is 
a positive advantage. 
For example, a crystal may provide an efficient and 
convenient way to store a protein.
Anecdotal evidence for this correlation between crystallizability 
and function can perhaps be found in the history of protein 
crystallization,\cite{McPherson99} 
as it is reasonable to expect that proteins that were among the
first to be crystallized are at the easier end of the 
spectrum of crystallizability.
For example, storage proteins, particularly the globulins found in 
seeds and nuts, were amongst the earlier protein crystals to 
be discovered, although this, at least partly, 
also reflects the ready availability of a protein source.

More direct evidence for this potential positive side to crystallization
comes from the identification of crystals {\it in vivo}, 
an interesting overview of which is given in Ref.\ \onlinecite{McPherson99}.  
For example, protein crystals have been observed in the egg yolks of 
various organisms, and ribosome crystals have been found 
in hibernating animals, presumably because they act as a temporary reservoir
for this important cellular component.
Particularly interesting in this regard is
the {\it Bacillus thuringiensis} class of bacteria, which produce 
protein toxins specific to a wide variety of insects.\cite{Schnepf98} 
Crystals provide a particularly stable (up to periods of years) 
form for these bacteria to store these toxins. When ingested,
these crystals dissolve, releasing the toxins
to attack the gut wall of the target insect, thus facilitating the 
entry of germinating bacterial spores into the host.

Although perhaps harmful to the host cell, there seems little
reason why the formation of crystals of virus particles
would be disadvantageous to the virus. Indeed, it probably 
presents a convenient way to densely pack the particles 
and so minimize possible constraints on self-replication.
Consistent with this supposition, crystals of spherical 
and icosahedral viruses are frequently observed in infected cells.
Furthermore, viruses were also amongst the earlier 
biological particles to be crystallized. 

Even more fascinating is the ingenious use of protein 
crystallization made by viruses that are able 
to form a quiescent state by embedding themselves
in a protein crystal matrix.\cite{Smith76} 
These viruses cause large quantities of an easily crystallizable protein 
to be expressed in an infected cell. 
Nucleation of crystals of this protein then occurs on 
the surface of the viral particles, surrounding them by crystal and
providing the viruses with a protective environment until further
transmission is possible. Similar to the bacterial toxins, these
crystals readily dissolve in the gut of the insect host, 
releasing the virus.

The important lesson from these examples 
is that when it is beneficial for the organism, nature seems
to have no difficulty enabling proteins to crystallize. 
Indeed, such crystals can form spontaneously in the cell simply
when the concentration is sufficiently high without the need 
for extremely high purities and a series of 
precipitants to drive the process.
The contrasting difficulty that most proteins have in crystallizing, therefore, 
does not seem to be an intrinsic property of 
polypeptide chains that have a well-defined folded structure. 
Rather, 
it is a property that has been selected by nature, because of 
the need for the protein-protein interactions to be 
strictly controlled if the cell is to function properly.

Our arguments are not undermined by the fact that proteins show a 
whole spectrum of crystallizabilities, with proteins
such as lysozymes, hemoglobins and insulins at the easier end. 
This is to be expected from our perspective. 
Firstly, as we have seen, the strength of the
selection pressure against crystallization may vary considerably (and 
even be reversed) depending on the function and environment experienced
by the protein. Secondly, evolution has no interest in controlling the
properties of proteins in non-physiological conditions, and so one should
not expect a uniform response. Instead, the degree to which the {\it in vivo}
low crystallizability carries over to {\it in vitro} environments
is likely to show significant variability.  Lastly, evolution 
just requires the crystallizability to be low enough to pose only
a low risk to the cell. 
But there is no reason why  the crystallizability could not 
be significantly below this threshold value, 
as long as it is not achieved at the expense of the other properties of 
the protein. 

Because the individual concentrations for the majority of
proteins are very low relative to the overall protein concentration,
some might argue that the putative negative design acts most directly 
against the non-specific aggregation of native proteins, and then,  
perhaps because the mechanisms used are generic, 
only indirectly against crystallization.
Indeed, the evidence that we have presented for
negative design with respect to crystallization 
does not indicate 
whether this effect is direct or indirect. Moreover, the
typical cellular concentration of a protein in the cell will be one of
the factors that determines the magnitude of the selection pressure
against crystallization.  However, it should also be remembered that
low concentrations do not prevent functional interactions between
proteins, and that the coexistence line between crystal and dilute
solution in a protein phase diagram can occur at very low
concentrations.\cite{Muschol97} In our opinion, the negative design against
crystallization is probably a mixture of direct and indirect effects.

In this article we have presented a different perspective by which to
rationalize the crystallizability of proteins.  Progress towards
enhancing the success rate of crystallizing proteins will depend on
unravelling the mechanisms by which nature achieves this negative
design.  We have highlighted several studies which show that random
mutations enhance crystallizability. Mutagenesis programs have already
led to important new insights into the nature of the functional
interactions between proteins\cite{Bogan98} and the key determinants
of the propensity for amyloidogenic aggregation.\cite{Chiti03}
Similar systematic studies may provide an important means for
understanding the mechanisms by which proteins are prevented from
crystallizing.  This would have the potential not only to provide
further confirmation of our negative design hypothesis, but also to
reveal residues and surface patterns that are key for the formation
or prevention of crystal contacts.

We have already highlighted some interesting results that flag the
potentially important role played by lsyine residues. Further, more
detailed physical studies of the mechanisms by which lysine influences
the protein-protein interactions would be desirable. For example, it 
would be interesting to see how the second virial coefficient, 
a measure of the strength of the generic attractions between proteins, 
changes with the mutation of surface lysine residues.
Computer simulations could also potentially provide a more 
detailed atomistic picture of the conformations adopted by a surface lysine
and how this changes with crystal contact formation.

Obtaining a better understanding of the
mechanisms used to hinder crystallization would open up the possibility of
finding ways to ``turn off'' these negative interactions, and so enhance
a protein's crystallizability. The required changes to the surface
properties could perhaps be achieved through mutations or the
addition of appropriate precipitants. Furthermore, such advances in
our understanding of protein crystallization could also potentially
rationalize the effects of some of the precipitants currently used. At
best, the effects of these precipitants are understood only in terms
of their effect on average properties, such as the second virial
coefficient. However, the mechanisms underlying some, e.g.\ polyethylene
glycol, remain rather mysterious.

Finally, we note that only positive outcomes of protein
crystallization experiments have traditionally been published.  In our
opinion, experiments where crystallizability is reduced rather than
enhanced may also contain useful information about the mechanisms of
negative design.  Thinking in terms of this principle may help
experimentalists decide when such ``negative'' results are
nevertheless valuable.

To summarize, we have presented a perspective on protein
crystallization whereby the difficulty crystallographers have in
obtaining protein crystals is a consequence of evolutionary negative
design against aggregation of native-state proteins.  It really is the
case that proteins do not want to crystallize because a protein that
is prone to crystallization, or in fact any form of aggregation, is
potentially deleterious to the cell.  The mechanisms of this negative
design are only very partially understood.
But our main point
is that 
understanding these mechanisms of {\em negative design} should provide
fruitful insights that lead to {\em positive} advances in 
crystallizing globular proteins.

\begin{acknowledgments}
The authors are grateful to the Royal Society for financial support,
and Luca Pellegrini for a careful reading of the manuscript.
The protein crystal image for the graphical table of contents was kindly
provided by Allan D'Arcy, Morphochem, Switzerland.
\end{acknowledgments}


\begin{thebibliography}{40}
\expandafter\ifx\csname natexlab\endcsname\relax\def\natexlab#1{#1}\fi
\expandafter\ifx\csname bibnamefont\endcsname\relax
  \def\bibnamefont#1{#1}\fi
\expandafter\ifx\csname bibfnamefont\endcsname\relax
  \def\bibfnamefont#1{#1}\fi
\expandafter\ifx\csname citenamefont\endcsname\relax
  \def\citenamefont#1{#1}\fi
\expandafter\ifx\csname url\endcsname\relax
  \def\url#1{\texttt{#1}}\fi
\expandafter\ifx\csname urlprefix\endcsname\relax\def\urlprefix{URL }\fi
\providecommand{\bibinfo}[2]{#2}
\providecommand{\eprint}[2][]{\url{#2}}

\bibitem[{\citenamefont{Bergfors}(1999)}]{Bergfors}
\bibinfo{editor}{\bibfnamefont{T.~M.} \bibnamefont{Bergfors}}, ed.,
  \emph{\bibinfo{title}{Protein Crystallization: Techniques, Strategies, and
  Tips}} (\bibinfo{publisher}{International University Line},
  \bibinfo{address}{La Jolla}, \bibinfo{year}{1999}).

\bibitem[{\citenamefont{Burley et~al.}(1999)\citenamefont{Burley, Almo,
  Bonanno, Capel, Chance, Gaasterland, Lin, Sali, Studier, and
  Swaminathan}}]{Burley99}
\bibinfo{author}{\bibfnamefont{S.~K.} \bibnamefont{Burley}},
  \bibinfo{author}{\bibfnamefont{S.~C.} \bibnamefont{Almo}},
  \bibinfo{author}{\bibfnamefont{J.~B.} \bibnamefont{Bonanno}},
  \bibinfo{author}{\bibfnamefont{M.}~\bibnamefont{Capel}},
  \bibinfo{author}{\bibfnamefont{M.~R.} \bibnamefont{Chance}},
  \bibinfo{author}{\bibfnamefont{T.}~\bibnamefont{Gaasterland}},
  \bibinfo{author}{\bibfnamefont{D.}~\bibnamefont{Lin}},
  \bibinfo{author}{\bibfnamefont{A.}~\bibnamefont{Sali}},
  \bibinfo{author}{\bibfnamefont{F.~W.} \bibnamefont{Studier}},
  \bibnamefont{and}
  \bibinfo{author}{\bibfnamefont{S.}~\bibnamefont{Swaminathan}},
  \bibinfo{journal}{Nat. Genet.} \textbf{\bibinfo{volume}{23}},
  \bibinfo{pages}{151} (\bibinfo{year}{1999}).

\bibitem[{\citenamefont{Vekilov and Chernov}(2002)}]{Vekilov02}
\bibinfo{author}{\bibfnamefont{P.~G.} \bibnamefont{Vekilov}} \bibnamefont{and}
  \bibinfo{author}{\bibfnamefont{A.~A.} \bibnamefont{Chernov}},
  \bibinfo{journal}{Solid State Phys.} \textbf{\bibinfo{volume}{57}},
  \bibinfo{pages}{1} (\bibinfo{year}{2002}).

\bibitem[{\citenamefont{Service}(2002)}]{Service02}
\bibinfo{author}{\bibfnamefont{R.~F.} \bibnamefont{Service}},
  \bibinfo{journal}{Science} \textbf{\bibinfo{volume}{298}},
  \bibinfo{pages}{948} (\bibinfo{year}{2002}).

\bibitem[{\citenamefont{Yee et~al.}(2003)\citenamefont{Yee, Pardee,
  Christendat, Savchenko, Edwards, and Arrowsmith}}]{Yee03}
\bibinfo{author}{\bibfnamefont{A.}~\bibnamefont{Yee}},
  \bibinfo{author}{\bibfnamefont{K.}~\bibnamefont{Pardee}},
  \bibinfo{author}{\bibfnamefont{D.}~\bibnamefont{Christendat}},
  \bibinfo{author}{\bibfnamefont{A.}~\bibnamefont{Savchenko}},
  \bibinfo{author}{\bibfnamefont{A.~M.} \bibnamefont{Edwards}},
  \bibnamefont{and} \bibinfo{author}{\bibfnamefont{C.~H.}
  \bibnamefont{Arrowsmith}}, \bibinfo{journal}{Acc. Chem. Res.}
  \textbf{\bibinfo{volume}{36}}, \bibinfo{pages}{183} (\bibinfo{year}{2003}).

\bibitem[{gla()}]{glass}
\bibinfo{note}{Glasses and gels most often form for systems where there is a
  network of strong bonds, eg SiO$_2$. By contrast, molecules with weak
  intermolecular interactions can relatively easily reorient to find the
  preferred orientation for the crystal.}

\bibitem[{\citenamefont{de~Gennes}(1979)}]{deGennes79}
\bibinfo{author}{\bibfnamefont{P.~G.} \bibnamefont{de~Gennes}},
  \emph{\bibinfo{title}{Scaling Concepts in Polymer Physics}}
  (\bibinfo{publisher}{Cornell University}, \bibinfo{address}{Ithaca},
  \bibinfo{year}{1979}).

\bibitem[{\citenamefont{Kopito}(2000)}]{Kopito00}
\bibinfo{author}{\bibfnamefont{R.~R.} \bibnamefont{Kopito}},
  \bibinfo{journal}{Trends Cell Biol.} \textbf{\bibinfo{volume}{10}},
  \bibinfo{pages}{524} (\bibinfo{year}{2000}).

\bibitem[{\citenamefont{Pande et~al.}(2001)\citenamefont{Pande, Pande, Asherie,
  Lomakin, Ogun, King, and Benedek}}]{Pande01}
\bibinfo{author}{\bibfnamefont{A.}~\bibnamefont{Pande}},
  \bibinfo{author}{\bibfnamefont{J.}~\bibnamefont{Pande}},
  \bibinfo{author}{\bibfnamefont{N.}~\bibnamefont{Asherie}},
  \bibinfo{author}{\bibfnamefont{A.}~\bibnamefont{Lomakin}},
  \bibinfo{author}{\bibfnamefont{O.}~\bibnamefont{Ogun}},
  \bibinfo{author}{\bibfnamefont{J.}~\bibnamefont{King}}, \bibnamefont{and}
  \bibinfo{author}{\bibfnamefont{G.~B.} \bibnamefont{Benedek}},
  \bibinfo{journal}{Proc. Natl. Acad. Sci. USA} \textbf{\bibinfo{volume}{98}},
  \bibinfo{pages}{6116} (\bibinfo{year}{2001}).

\bibitem[{\citenamefont{Vekilov et~al.}(2002)\citenamefont{Vekilov,
  Feeling-Taylor, Petsev, Galkin, Nagel, and Hirsch}}]{Vekilov02b}
\bibinfo{author}{\bibfnamefont{P.~G.} \bibnamefont{Vekilov}},
  \bibinfo{author}{\bibfnamefont{A.~R.} \bibnamefont{Feeling-Taylor}},
  \bibinfo{author}{\bibfnamefont{D.~N.} \bibnamefont{Petsev}},
  \bibinfo{author}{\bibfnamefont{O.}~\bibnamefont{Galkin}},
  \bibinfo{author}{\bibfnamefont{R.~L.} \bibnamefont{Nagel}}, \bibnamefont{and}
  \bibinfo{author}{\bibfnamefont{R.~E.} \bibnamefont{Hirsch}},
  \bibinfo{journal}{Biophys. J.} \textbf{\bibinfo{volume}{83}},
  \bibinfo{pages}{1147} (\bibinfo{year}{2002}).

\bibitem[{\citenamefont{McPherson}(1999)}]{McPherson99}
\bibinfo{author}{\bibfnamefont{A.}~\bibnamefont{McPherson}},
  \emph{\bibinfo{title}{Crystallization of biological macromolecules}}
  (\bibinfo{publisher}{Cold Spring Harbour Laboratory Press},
  \bibinfo{address}{Cold Spring Harbour}, \bibinfo{year}{1999}).

\bibitem[{\citenamefont{Hellinga}(1997)}]{Hellinga97}
\bibinfo{author}{\bibfnamefont{H.~W.} \bibnamefont{Hellinga}},
  \bibinfo{journal}{Proc. Natl. Acad. Sci. USA} \textbf{\bibinfo{volume}{94}},
  \bibinfo{pages}{10015} (\bibinfo{year}{1997}).

\bibitem[{\citenamefont{DeGrado et~al.}(1999)\citenamefont{DeGrado, Summa,
  Pavone, Nastri, and Lombardi}}]{DeGrado99}
\bibinfo{author}{\bibfnamefont{W.~F.} \bibnamefont{DeGrado}},
  \bibinfo{author}{\bibfnamefont{C.~M.} \bibnamefont{Summa}},
  \bibinfo{author}{\bibfnamefont{V.}~\bibnamefont{Pavone}},
  \bibinfo{author}{\bibfnamefont{F.}~\bibnamefont{Nastri}}, \bibnamefont{and}
  \bibinfo{author}{\bibfnamefont{A.}~\bibnamefont{Lombardi}},
  \bibinfo{journal}{Annu. Rev. Biochem.} \textbf{\bibinfo{volume}{68}},
  \bibinfo{pages}{779} (\bibinfo{year}{1999}).

\bibitem[{\citenamefont{Bolon}(2001)}]{Bolon01}
\bibinfo{author}{\bibfnamefont{D.~N.} \bibnamefont{Bolon}},
  \bibinfo{journal}{Biochemistry} \textbf{\bibinfo{volume}{40}},
  \bibinfo{pages}{10047} (\bibinfo{year}{2001}).

\bibitem[{\citenamefont{Richardson and Richardson}(2002)}]{Richardson02}
\bibinfo{author}{\bibfnamefont{J.~S.} \bibnamefont{Richardson}}
  \bibnamefont{and} \bibinfo{author}{\bibfnamefont{D.~C.}
  \bibnamefont{Richardson}}, \bibinfo{journal}{Proc. Natl. Acad. Sci. USA}
  \textbf{\bibinfo{volume}{99}}, \bibinfo{pages}{2754} (\bibinfo{year}{2002}).

\bibitem[{\citenamefont{Wang and Hecht}(2002)}]{Wang02b}
\bibinfo{author}{\bibfnamefont{W.}~\bibnamefont{Wang}} \bibnamefont{and}
  \bibinfo{author}{\bibfnamefont{M.~H.} \bibnamefont{Hecht}},
  \bibinfo{journal}{Proc. Natl. Acad. Sci. USA} \textbf{\bibinfo{volume}{99}},
  \bibinfo{pages}{2760} (\bibinfo{year}{2002}).

\bibitem[{\citenamefont{Havranek and Harbury}(2003)}]{Havranek03}
\bibinfo{author}{\bibfnamefont{J.~J.} \bibnamefont{Havranek}} \bibnamefont{and}
  \bibinfo{author}{\bibfnamefont{P.~B.} \bibnamefont{Harbury}},
  \bibinfo{journal}{Nat. Struct. Biol.} \textbf{\bibinfo{volume}{10}},
  \bibinfo{pages}{45} (\bibinfo{year}{2003}).

\bibitem[{\citenamefont{Shifman and Mayo}(2003)}]{Shifman03}
\bibinfo{author}{\bibfnamefont{J.~M.} \bibnamefont{Shifman}} \bibnamefont{and}
  \bibinfo{author}{\bibfnamefont{S.~L.} \bibnamefont{Mayo}},
  \bibinfo{journal}{Proc. Natl. Acad. Sci. USA} \textbf{\bibinfo{volume}{100}},
  \bibinfo{pages}{13274} (\bibinfo{year}{2003}).

\bibitem[{\citenamefont{Ellis and Minton}(2003)}]{Ellis03}
\bibinfo{author}{\bibfnamefont{R.~J.} \bibnamefont{Ellis}} \bibnamefont{and}
  \bibinfo{author}{\bibfnamefont{A.~P.} \bibnamefont{Minton}},
  \bibinfo{journal}{Nature} \textbf{\bibinfo{volume}{425}}, \bibinfo{pages}{27}
  (\bibinfo{year}{2003}).

\bibitem[{\citenamefont{Sear}(2004)}]{Sear03b}
\bibinfo{author}{\bibfnamefont{R.~P.} \bibnamefont{Sear}}, \bibinfo{journal}{J.
  Chem. Phys.} \textbf{\bibinfo{volume}{120}}, \bibinfo{pages}{998}
  (\bibinfo{year}{2004}).

\bibitem[{Sea()}]{Searunpub}
\bibinfo{note}{R. P. Sear, unpublished}.

\bibitem[{\citenamefont{Bastolla et~al.}(2003)\citenamefont{Bastolla, Porto,
  Roman, and Vendruscolo}}]{Bastolla03}
\bibinfo{author}{\bibfnamefont{U.}~\bibnamefont{Bastolla}},
  \bibinfo{author}{\bibfnamefont{M.}~\bibnamefont{Porto}},
  \bibinfo{author}{\bibfnamefont{H.~E.} \bibnamefont{Roman}}, \bibnamefont{and}
  \bibinfo{author}{\bibfnamefont{M.}~\bibnamefont{Vendruscolo}},
  \bibinfo{journal}{J. Mol. Evol.} \textbf{\bibinfo{volume}{56}},
  \bibinfo{pages}{243} (\bibinfo{year}{2003}).

\bibitem[{\citenamefont{McElroy et~al.}(1992)\citenamefont{McElroy, Sisson,
  Schoettlin, Aust, and Villafranca}}]{McElroy92}
\bibinfo{author}{\bibfnamefont{H.~E.} \bibnamefont{McElroy}},
  \bibinfo{author}{\bibfnamefont{G.~W.} \bibnamefont{Sisson}},
  \bibinfo{author}{\bibfnamefont{W.~E.} \bibnamefont{Schoettlin}},
  \bibinfo{author}{\bibfnamefont{R.~M.} \bibnamefont{Aust}}, \bibnamefont{and}
  \bibinfo{author}{\bibfnamefont{J.~E.} \bibnamefont{Villafranca}},
  \bibinfo{journal}{J. Cryst. Growth} \textbf{\bibinfo{volume}{122}},
  \bibinfo{pages}{265} (\bibinfo{year}{1992}).

\bibitem[{\citenamefont{D'Arcy et~al.}(1999)\citenamefont{D'Arcy, Stihle,
  Kostrewa, and Dale}}]{Darcy99}
\bibinfo{author}{\bibfnamefont{A.}~\bibnamefont{D'Arcy}},
  \bibinfo{author}{\bibfnamefont{M.}~\bibnamefont{Stihle}},
  \bibinfo{author}{\bibfnamefont{D.}~\bibnamefont{Kostrewa}}, \bibnamefont{and}
  \bibinfo{author}{\bibfnamefont{G.}~\bibnamefont{Dale}},
  \bibinfo{journal}{Acta Crystallogr. D} \textbf{\bibinfo{volume}{55}},
  \bibinfo{pages}{1623} (\bibinfo{year}{1999}).

\bibitem[{\citenamefont{Dale et~al.}(2003)\citenamefont{Dale, Oefner, and
  D'Arcy}}]{Dale03}
\bibinfo{author}{\bibfnamefont{G.~E.} \bibnamefont{Dale}},
  \bibinfo{author}{\bibfnamefont{C.}~\bibnamefont{Oefner}}, \bibnamefont{and}
  \bibinfo{author}{\bibfnamefont{A.}~\bibnamefont{D'Arcy}},
  \bibinfo{journal}{J. Struct. Biol.} \textbf{\bibinfo{volume}{142}},
  \bibinfo{pages}{88} (\bibinfo{year}{2003}).

\bibitem[{\citenamefont{Bogan and Thorn}(1998)}]{Bogan98}
\bibinfo{author}{\bibfnamefont{A.~A.} \bibnamefont{Bogan}} \bibnamefont{and}
  \bibinfo{author}{\bibfnamefont{K.~S.} \bibnamefont{Thorn}},
  \bibinfo{journal}{J. Mol. Biol.} \textbf{\bibinfo{volume}{280}},
  \bibinfo{pages}{1} (\bibinfo{year}{1998}).

\bibitem[{\citenamefont{Baud and Karlin}(1999)}]{Baud99}
\bibinfo{author}{\bibfnamefont{F.}~\bibnamefont{Baud}} \bibnamefont{and}
  \bibinfo{author}{\bibfnamefont{S.}~\bibnamefont{Karlin}},
  \bibinfo{journal}{Proc. Natl. Acad. Sci. USA} \textbf{\bibinfo{volume}{96}},
  \bibinfo{pages}{12494} (\bibinfo{year}{1999}).

\bibitem[{\citenamefont{Lins et~al.}(2003)\citenamefont{Lins, Thomas, and
  Brasseur}}]{Lins03}
\bibinfo{author}{\bibfnamefont{L.}~\bibnamefont{Lins}},
  \bibinfo{author}{\bibfnamefont{A.}~\bibnamefont{Thomas}}, \bibnamefont{and}
  \bibinfo{author}{\bibfnamefont{R.}~\bibnamefont{Brasseur}},
  \bibinfo{journal}{Protein Sci.} \textbf{\bibinfo{volume}{12}},
  \bibinfo{pages}{1406} (\bibinfo{year}{2003}).

\bibitem[{\citenamefont{Dasgupta et~al.}(1997)\citenamefont{Dasgupta, Iyer,
  Bryant, Lawrence, and Bell}}]{Dasgupta97}
\bibinfo{author}{\bibfnamefont{S.}~\bibnamefont{Dasgupta}},
  \bibinfo{author}{\bibfnamefont{G.~H.} \bibnamefont{Iyer}},
  \bibinfo{author}{\bibfnamefont{S.~H.} \bibnamefont{Bryant}},
  \bibinfo{author}{\bibfnamefont{C.~E.} \bibnamefont{Lawrence}},
  \bibnamefont{and} \bibinfo{author}{\bibfnamefont{J.~A.} \bibnamefont{Bell}},
  \bibinfo{journal}{Proteins} \textbf{\bibinfo{volume}{28}},
  \bibinfo{pages}{494} (\bibinfo{year}{1997}).

\bibitem[{\citenamefont{Iyer et~al.}(2000)\citenamefont{Iyer, Dasgupta, and
  Bell}}]{Iyer00}
\bibinfo{author}{\bibfnamefont{G.~H.} \bibnamefont{Iyer}},
  \bibinfo{author}{\bibfnamefont{S.}~\bibnamefont{Dasgupta}}, \bibnamefont{and}
  \bibinfo{author}{\bibfnamefont{J.~A.} \bibnamefont{Bell}},
  \bibinfo{journal}{J. Cryst. Growth} \textbf{\bibinfo{volume}{217}},
  \bibinfo{pages}{429} (\bibinfo{year}{2000}).

\bibitem[{\citenamefont{Lo~Conte et~al.}(1999)\citenamefont{Lo~Conte, Chothia,
  and Janin}}]{LoConte99}
\bibinfo{author}{\bibfnamefont{L.}~\bibnamefont{Lo~Conte}},
  \bibinfo{author}{\bibfnamefont{C.}~\bibnamefont{Chothia}}, \bibnamefont{and}
  \bibinfo{author}{\bibfnamefont{J.}~\bibnamefont{Janin}}, \bibinfo{journal}{J.
  Mol. Biol.} \textbf{\bibinfo{volume}{285}}, \bibinfo{pages}{2177}
  (\bibinfo{year}{1999}).

\bibitem[{\citenamefont{Jones and Thornton}(1996)}]{Jones96}
\bibinfo{author}{\bibfnamefont{S.}~\bibnamefont{Jones}} \bibnamefont{and}
  \bibinfo{author}{\bibfnamefont{J.~M.} \bibnamefont{Thornton}},
  \bibinfo{journal}{Proc. Natl. Acad. Sci. USA} \textbf{\bibinfo{volume}{93}},
  \bibinfo{pages}{13} (\bibinfo{year}{1996}).

\bibitem[{\citenamefont{Longenecker
  et~al.}(2001{\natexlab{a}})\citenamefont{Longenecker, Garrard, Sheffield, and
  Derewenda}}]{Longenecker01}
\bibinfo{author}{\bibfnamefont{K.~L.} \bibnamefont{Longenecker}},
  \bibinfo{author}{\bibfnamefont{S.~M.} \bibnamefont{Garrard}},
  \bibinfo{author}{\bibfnamefont{P.~J.} \bibnamefont{Sheffield}},
  \bibnamefont{and} \bibinfo{author}{\bibfnamefont{Z.~S.}
  \bibnamefont{Derewenda}}, \bibinfo{journal}{Acta Crystallogr. D}
  \textbf{\bibinfo{volume}{57}}, \bibinfo{pages}{679}
  (\bibinfo{year}{2001}{\natexlab{a}}).

\bibitem[{\citenamefont{Longenecker
  et~al.}(2001{\natexlab{b}})\citenamefont{Longenecker, Lewis, Chikumi,
  Gutkind, and Derewenda}}]{Longenecker01b}
\bibinfo{author}{\bibfnamefont{K.~L.} \bibnamefont{Longenecker}},
  \bibinfo{author}{\bibfnamefont{M.~E.} \bibnamefont{Lewis}},
  \bibinfo{author}{\bibfnamefont{H.}~\bibnamefont{Chikumi}},
  \bibinfo{author}{\bibfnamefont{J.~S.} \bibnamefont{Gutkind}},
  \bibnamefont{and} \bibinfo{author}{\bibfnamefont{Z.~S.}
  \bibnamefont{Derewenda}}, \bibinfo{journal}{Structure}
  \textbf{\bibinfo{volume}{3}}, \bibinfo{pages}{559}
  (\bibinfo{year}{2001}{\natexlab{b}}).

\bibitem[{\citenamefont{Garrard et~al.}(2001)\citenamefont{Garrard,
  Longenecker, Lewis, Sheffield, and Derewenda}}]{Garrard01}
\bibinfo{author}{\bibfnamefont{S.~M.} \bibnamefont{Garrard}},
  \bibinfo{author}{\bibfnamefont{K.~L.} \bibnamefont{Longenecker}},
  \bibinfo{author}{\bibfnamefont{M.~E.} \bibnamefont{Lewis}},
  \bibinfo{author}{\bibfnamefont{P.~J.} \bibnamefont{Sheffield}},
  \bibnamefont{and} \bibinfo{author}{\bibfnamefont{Z.~S.}
  \bibnamefont{Derewenda}}, \bibinfo{journal}{Protein Expres. Purif.}
  \textbf{\bibinfo{volume}{21}}, \bibinfo{pages}{412} (\bibinfo{year}{2001}).

\bibitem[{\citenamefont{Mateja et~al.}(2002)\citenamefont{Mateja, Devdjiev,
  Krowarsch, Longenecker, Dauter, Otlewski, and Derewenda}}]{Mateja02}
\bibinfo{author}{\bibfnamefont{A.}~\bibnamefont{Mateja}},
  \bibinfo{author}{\bibfnamefont{Y.}~\bibnamefont{Devdjiev}},
  \bibinfo{author}{\bibfnamefont{D.}~\bibnamefont{Krowarsch}},
  \bibinfo{author}{\bibfnamefont{K.~L.} \bibnamefont{Longenecker}},
  \bibinfo{author}{\bibfnamefont{Z.}~\bibnamefont{Dauter}},
  \bibinfo{author}{\bibfnamefont{J.}~\bibnamefont{Otlewski}}, \bibnamefont{and}
  \bibinfo{author}{\bibfnamefont{Z.~S.} \bibnamefont{Derewenda}},
  \bibinfo{journal}{Acta Crystallogr. D} \textbf{\bibinfo{volume}{58}},
  \bibinfo{pages}{1983} (\bibinfo{year}{2002}).

\bibitem[{\citenamefont{Schnepf et~al.}(1998)\citenamefont{Schnepf, Crickmore,
  Van~Rie, Lereclus, Baum, Feitelson, Zeigler, and H.}}]{Schnepf98}
\bibinfo{author}{\bibfnamefont{E.}~\bibnamefont{Schnepf}},
  \bibinfo{author}{\bibfnamefont{N.}~\bibnamefont{Crickmore}},
  \bibinfo{author}{\bibfnamefont{J.}~\bibnamefont{Van~Rie}},
  \bibinfo{author}{\bibfnamefont{D.}~\bibnamefont{Lereclus}},
  \bibinfo{author}{\bibfnamefont{J.}~\bibnamefont{Baum}},
  \bibinfo{author}{\bibfnamefont{J.}~\bibnamefont{Feitelson}},
  \bibinfo{author}{\bibfnamefont{D.~R.} \bibnamefont{Zeigler}},
  \bibnamefont{and} \bibinfo{author}{\bibfnamefont{D.~D.} \bibnamefont{H.}},
  \bibinfo{journal}{Microbiol. Mol. Biol. Rev.} \textbf{\bibinfo{volume}{62}},
  \bibinfo{pages}{775} (\bibinfo{year}{1998}).

\bibitem[{\citenamefont{Smith}(1976)}]{Smith76}
\bibinfo{author}{\bibfnamefont{K.~M.} \bibnamefont{Smith}},
  \emph{\bibinfo{title}{Virus-Insect Relationships}}
  (\bibinfo{publisher}{Longman}, \bibinfo{address}{London},
  \bibinfo{year}{1976}).

\bibitem[{\citenamefont{Muschol and Rosenberger}(1997)}]{Muschol97}
\bibinfo{author}{\bibfnamefont{M.}~\bibnamefont{Muschol}} \bibnamefont{and}
  \bibinfo{author}{\bibfnamefont{F.}~\bibnamefont{Rosenberger}},
  \bibinfo{journal}{J. Chem. Phys.} \textbf{\bibinfo{volume}{64}},
  \bibinfo{pages}{1953} (\bibinfo{year}{1997}).

\bibitem[{\citenamefont{Chiti et~al.}(2003)\citenamefont{Chiti, Stefani,
  Taddei, Ramponi, and Dobson}}]{Chiti03}
\bibinfo{author}{\bibfnamefont{F.}~\bibnamefont{Chiti}},
  \bibinfo{author}{\bibfnamefont{M.}~\bibnamefont{Stefani}},
  \bibinfo{author}{\bibfnamefont{N.}~\bibnamefont{Taddei}},
  \bibinfo{author}{\bibfnamefont{G.}~\bibnamefont{Ramponi}}, \bibnamefont{and}
  \bibinfo{author}{\bibfnamefont{C.~M.} \bibnamefont{Dobson}},
  \bibinfo{journal}{Nature} \textbf{\bibinfo{volume}{424}},
  \bibinfo{pages}{805} (\bibinfo{year}{2003}).

\end{thebibliography}
\end{document}